\documentclass[letterpaper,twocolumn,10pt]{article}
\usepackage{usenix,epsfig,endnotes}

\usepackage{epstopdf}%
\usepackage{amsmath}%
\usepackage{amsfonts}%
\usepackage{amssymb}%
\usepackage{graphicx}
\usepackage{wrapfig}
\usepackage{subfigure}
\usepackage{algorithm2e}
\usepackage[all]{xy}
\usepackage{sidecap}
\usepackage{url}

\newcommand{\wini}{\mbox{$\mathsf{C}$}}%   {$\textit{Wini}$}
\newcommand{\lin}{\mbox{$\mathsf{S}$}}%{$\textit{Lin}$}
\newcommand{\mal}{\mbox{$\mathsf{Mallory}$}}%{$\textit{Mal}$}
\newcommand{\ests}{\textit{$e_{\textit{wnd}}$}}
\newcommand{\esta}{\textit{$e_{\textit{una}}$}}
\newcommand{\cwnd}{\textit{wnd}}

\newcommand{\scom}{\mbox{$\mathsf{www.s.com}$}}
\newcommand{\malcom}{\mbox{$\mathsf{www.mallory.com}$}}%{$\textit{Mal}$}
\newcommand{\victimserv}{\mbox{$\mathsf{www.victim\text{\textendash}server.com}$}}%{$\textit{Mal}$}

\newcommand{\ignore}[1]{}

\begin{document}

\date{}
\author{Yossi Gilad and Amir Herzberg\\Department of Computer Science Bar Ilan University}

\title{Off-Path Attacking the Web}
%Blindly Injecting Data into TCP Connections\\
%{\small \it ``TCP was never designed to be a secure protocol'' -- Steven M. Bellovin \cite{Bellovin:Look:Back:at:TCP:IP:Security}}\\
%{\small \it ``Security is, I would say, our top priority'' -- Bill Gates \cite{Gates:Security}}

\maketitle

%\today

\begin{abstract}
We show how an off-path (spoofing-only) attacker can 
%create a long-lived TCP connection between a victim client that enters his website to another web-server. We show how the attacker can then 
perform cross-site scripting (XSS), cross-site request forgery (CSRF) and site spoofing/defacement attacks, 
without requiring vulnerabilities in either web-browser or server and circumventing known defenses. Attacker can also launch devastating denial of service (DoS) attacks,
%against the client, server and even the Internet core. The DoS attacks can be launched 
even when the connection between the client and the server is secured with SSL/TLS.
The attacks are practical and require a puppet (malicious script in browser sandbox) running on a the victim client machine, and attacker capable of IP-spoofing on the Internet. 

Our attacks use a technique allowing an off-path attacker to learn the sequence numbers of both client and server in a TCP connection. 
%We build on an existing technique \cite{lkm:phrack:07} that allows such off-path attacker to learn both sequence numbers of a TCP connection. 
The technique exploits the fact that many computers, in particular those running Windows, use a global IP-ID counter, which provides a side channel allowing efficient exposure of the connection sequence numbers. %, (still) available in all Windows versions, to probe the client machine. 

We present results of experiments evaluating the learning technique and the attacks that exploit it. Finally, we present practical defenses that can be deployed at the firewall level; no changes to existing TCP/IP stacks are required.

%While most of the specific flaws exploited by this attack may (and should) be fixed, the implications are long-term and should motivate adoption of cryptography, especially IPsec. 

\end{abstract}
\section{Introduction} \label{intro}
TCP is the main transport protocol over the Internet, ensuring reliable and efficient connections. TCP was {\em not} designed to be secure against {\em Man-in-the-Middle (MitM)}; in fact, it is trivially vulnerable to MitM attacks. However, it seems that man-in-the-middle and eavesdropping attacks are relatively rare in practice, since they require the attacker to control routers or links {\em along the path} between the victims. 
Instead, most practical attacks involve malicious hosts, without MitM capabilities, i.e., the attackers are {\em off-path}. 

In our attacks, as well as in many other off-path attacks (e.g., SYN-flood \cite{rfc4953}, DNS-poisoning \cite{kaminsky:dns}), the attacker sends {\em spoofed packets}, i.e., packets with fake (spoofed) sender IP address. Due to ingress filtering \cite{rfc3013,rfc2827,rfc3704} and other anti-spoofing measures, IP spoofing is less commonly available than before, but still feasible, see \cite{SpooferProject,%pang2004cib,beverly2005spi,
journals/toit/EhrenkranzL09}. Apparently, there is still a significant number of ISPs that do not perform ingress filtering for their clients (especially to multihomed customers). Furthermore,
with the growing concern of cyberwarfare and cybercrime, some ISPs may intentionally support spoofing. Hence, it is still reasonable to assume spoofing ability. 

However, there is a widespread belief that an `off-path' spoofing attacker, cannot {\em inject} traffic into a TCP connection. The reasoning is that an incoming TCP packet must contain valid sequence number (or be discarded); the sequence number field is $32$ bits long and initialized using randomness, therefore, it seems unlikely that an attacker can efficiently generate a spoofed packet which will be accepted by the recipient, i.e., inject data into the TCP stream. 

%This belief is demonstrated in the limited adoption of security mechanisms such as IPsec \cite{rfc4301}, SSL/TLS \cite{rfc5246} and DNSsec \cite{rfc4033}. The 
This belief is also stated in RFCs and standards, e.g., in RFC 4953, discussing on TCP spoofing attacks (see \cite{rfc4953}, Section 2.2). 
Indeed, since its early days, %and until now, 
most Internet traffic uses TCP - and is not cryptographically protected, in spite of warnings, e.g., by Morris \cite{Morris85}, Bellovin \cite{Bellovin:security:problems:in:TCP,Bellovin:Look:Back:at:TCP:IP:Security}). 

Of course, 
TCP injections are easy, for implementations using predictable initial sequence numbers (ISNs). This was observed already by Morris at 1985 \cite{Morris85} and abused %a decade later 
by Mitnick \cite{Shimomura:1995:TPC:546748}. 
% \cite{Morris85,morris1,Shimomura:1995:TPC:546748,Tsutomu}. 
Later, at 2001, Zalewski found that most implementations still used predictable sequence numbers \cite{zalewski2001strange}. 

However, by now, most or all major implementations ensure sufficiently-unpredictable initial sequence numbers, e.g., following \cite{rfc6528,Gont11TCP}. 
Does this imply that TCP injections are infeasible? 

Zalewski \cite{Zalewski} suggested that it may be possible to spoof a non-first fragment of a (fragmented) TCP packet, when the values of the fragment IP-IDs are predictable. In particular,  existing implementations of Windows use {\em globally-incrementing IP-ID values}, which are easy to predict. Zalewski's attack may also be applicable to Linux, using the IP-ID prediction techniques in \cite{GH:WOOT11}. However, exploiting such non-first-fragment TCP injections seems challenging; furthermore, currently almost all TCP implementations use  path MTU discovery  \cite{rfc1191,rfc4821} and avoid fragmentation completely. Hence, Zalewski's attack (on TCP) will rarely work in practice. 

%In this paper, we investigate TCP injections assuming the use of unpredictable ISNs. Our results are related to two informal proposals for TCP injections (with unpredictable ISNs), each of which following a separate approach were presented in the recent years. 
%
%------------

Yet, we show that TCP injections are still possible. We present an efficient, practical technique, based on globally-incrementing IP-ID, allowing an off-path adversary, \mal, to inject data into a TCP connection between two communicating peers: a client \wini\ and a server \lin. The attack is not immediate, and requires a connection lasting a few dozens of seconds. We present experimental results, showing that our techniques allow efficient, practical TCP injections. Furthermore, we show, that the attacks have significant potential for abuse. Specifically, we show how our TCP injection techniques, allow both {\em circumvention of the Same Origin Policy} \cite{rfc6454,Zalewski:web:sec:book} and {\em devastating DoS attacks}. Details follow.

%In \cite{klm:phrack:07}, klm reported a more alarming abuse of globally-incrementing IP-ID values, allowing injection of traffic into TCP connections; that attack had some limitations (e.g., bypassing of firewalls). 

Our TCP injection technique is related to a proposal for TCP injection, by klm \cite{lkm:phrack:07}. The technique described by klm had some limitations, e.g., it did not work for clients connected by a firewall. More significantly, klm did not present experimental results; our experiments show that their technique, even with some improvements, results in low injection success rates, unless the attacker has low latency to the victim (as when they are on the same LAN). Our techniques avoid this limitation. We provide a detailed comparison between our technique to \cite{lkm:phrack:07} in appendix \ref{appendix:comparison}.

Like \cite{Zalewski,GH:WOOT11,lkm:phrack:07}, our technique is based on the  predictability of the  IP-ID (e.g., in Windows); however, instead of using the predictable IP-ID to intercept or modify fragmented packets, as in \cite{Zalewski,GH:WOOT11}, we use the changes in the IP-ID as a {\em side channel}. 
We use this side-channel to allow the attacker   
to detect difference in responses for crafted probe packets that she sends to the client; our implementation uses specific probes, but others may exist.  
\ignore{This allows the attacker to learn whether the client had sent a packet in response to the attacker's probe packets, by observing the IP-ID in packets that the attacker receives from the client. }

Previous works noted that the predictable IP-ID can be used as a side channel, 
%In particular, in \cite{rfc6274}, Gont explains that 
allowing an attacker to use one connection to learn about events in another connection, which is undesirable.
Gont \cite{rfc6274} mentions several ways in which the side channel based on globally-incremented IP-ID can be abused. However, their impact is modest. 
In particular, the side-channel can be used to perform the idle (stealth) scan attack \cite{zalewski2005silence,Klein07:OpenBSD,nmap:book}, and to count the number of machines behind NAT \cite{conf/imc/Bellovin02}. 

%In many systems, including most or all versions of Windows, the IP-ID is globally-incrementing. Namely, the IP-ID is simply incremented for every packet sent. 
%As Gont explains in \cite{rfc6274}, this creates a side-channel, i.e., attacker can use one connection to learn about events in another connection, which is undesirable. 

%Our technique exploits a side-channel provided by {\em globally-incrementing IP-ID}. In IPv4, the IP-ID is a 16-bit identifier included in each IP header; in IPv6, it is 32 bits. All fragments of a fragmented packet must use the same IP-ID, to allow defragmentation. Hence, the IP-ID be distinct for fragments of different packets (with same source, destination and protocol). 

%We present this technique in Section \ref{Injection:SeqExposure}.

However, vendors continued using globally-incrementing IP-ID values, even after we presented our initial TCP injection results to them,\ignore{ \cite{ms:disc:march11},} and being aware of the previous attacks exploiting the globally incrementing IP-IDs. Their justification is that they believed that such attacks are impractical and too complex to be exploited in practice. They confirmed our results and agreed that they are feasible; they will address the problem in new releases. 

We believe that this response is `too little, too late', and that it is critical for the community to be aware of the threat and apply mitigations (Section \ref{Injection:Defenses}). A more controversial conclusion is the need to apply more prudent approach to network security vulnerabilities, and respond to early indications of weakness, without waiting for a complete exploit; see discussion in Section \ref{Injection:Conclusions}. 

 %We next briefly discuss each of these two issues. 

Much of our work focused on analyzing what are the practical implications of the TCP injection.
We study two approaches to exploit TCP injections: to circumvent {\em address based server authentication}, usually referred to as {\em Same Origin Policy (SOP)} \cite{rfc6454,Zalewski:web:sec:book}; and to launch a {\em Denial of Service (DoS)} attack. We discuss each of these briefly in separate subsections below, and in depth in Sections \ref{exploiting:S} and \ref{exploits:DoS}. All attacks are based on the same attacker and network assumptions which we now describe. 

\subsubsection*{Attacker Capabilities and Network Model} \label{intro:model}

\begin{figure}
  \begin{center}
    \includegraphics[width=0.35\textwidth]{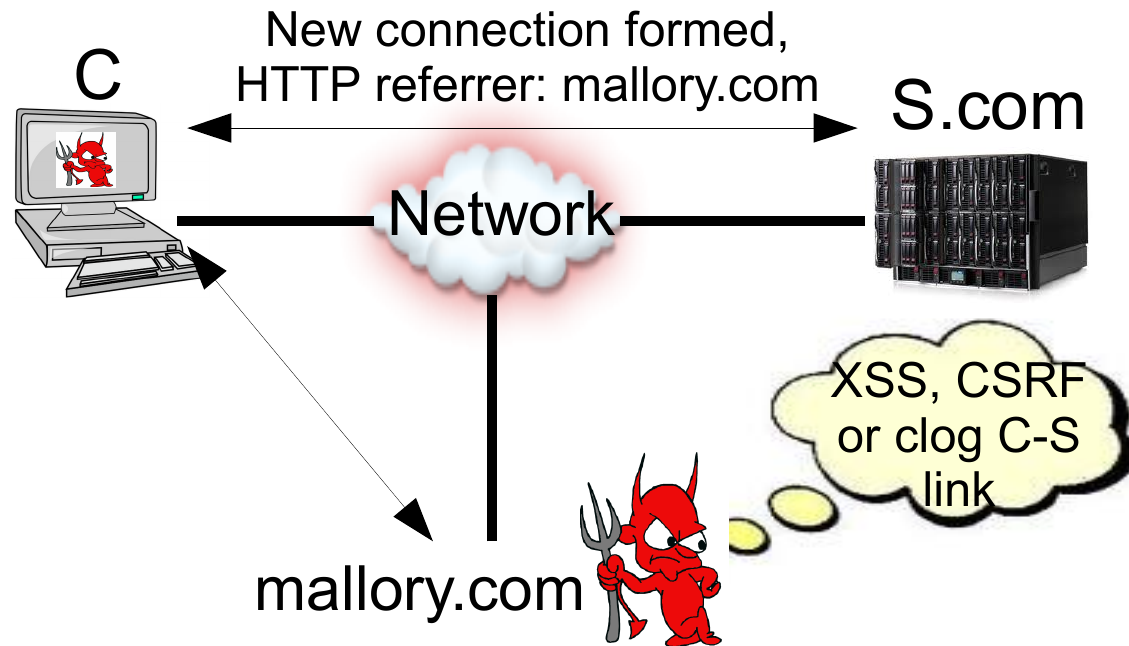}
  \end{center}
  \label{fig:ourmodel}
   \vspace{-15pt}
  \caption{Network Model. \wini\ enters \malcom, the adversarial web page. A script on that page forms a connection with \scom.}
\end{figure}

All our attacks work in the same settings: an off-path, IP-spoofing attacker. We also assume that the attacker is  able to control some {\em puppets} \cite{AATA08:puppetnets}, i.e., scripts, applets or other restricted (sandboxed) programs, running on client machines accessing an adversarial web site. This is illustrated in Figure \ref{fig:ourmodel} where \wini\ enters a site controlled by the adversary, \mal. This allows \mal\ to run a malicious script within \wini's browser sandbox. The script allows \mal\ to (1) {\em form} the connection between \wini\ and \lin, and (2) probe \wini's connection with \lin\ and avoid firewall filtering. The first allows \mal\ to choose the victim server (\lin), we show how the second allows exposure of the TCP connection's four tuple. 

As mentioned above, our attacks require that the connection from \wini\ and \lin\ will not be short; since \mal\ forms this connection (using the puppet), she can ensure that it does not terminate by sending periodic requests to the server. In persistent HTTP connections, all requests are over the same (victim) connection and ensure it does not close. Persistent HTTP connections are the default configuration of apache servers and are also employed by many large web-servers (e.g., Facebook, Yahoo!, Google), but not all (e.g., live.com, the Japanese version of Yahoo!). Our attacks are browser independent, as we illustrate in experiments in the following sections.

\subsection{Breaking SOP and Address-Based Authentication}
TCP injection attacks were key to some of the most well known exploits, specifically, attacks against {\em address based client authentication}, e.g., see \cite{Shimomura:1995:TPC:546748,Bellovin:Look:Back:at:TCP:IP:Security}. However, as a result, address-based client authentication has become essentially obsolete, and mostly replaced with secure alternatives such as SSH and SSL/TLS. We believe that the only widely-deployed use of address based client authentication, is to identify clients involved in DoS attacks such as SYN flooding; and this threat can be dealt with by simple client-response authentication, possibly using cookies to avoid state-exhaustion on the server \cite{rfc4987}.  

However, current web security still relies, to large extent, on the {\em Same Origin Policy} \cite{rfc6454,Zalewski:web:sec:book}, i.e., on {\em address based server authentication}; our results show that relying on addresses to authenticate the {\em servers} is also risky. 

Using TCP injections to attack address based server authentication, e.g., to perform XSS attacks, is more challenging than using it to attack address based client authentication. In attacks on address based {\em client} authentication, the off-path attacker sends the initial SYN to open a new connection; hence, she knows the client's sequence number, as well as the source and destination IP addresses and ports; she `only' needs to predict the server's sequence number. In contrast, to attack address based {\em server} authentication, the off-path attacker  must guess {\em both} sequence numbers, as well as the IP addresses and ports of both parties; guessing the client port could also be challenging, as the client may choose it arbitrarily.% (even randomly).  

%\subsection{Contributions}
%\label{Into:Contributions}

%All attacks assume (only) an off-path, spoofing attacker, controlling (only) a {\em puppet} (sandboxed malware, see \cite{AATA08:puppetnets}) on the victim client machine.

%\subsection{The Challenges in Performing Off-Path TCP Injection Attacks}
%
%There are two challenges, the first is to allow an off-path attacker to inject arbitrary data into a TCP session. 
%For this task the adversary must (1) identify the existence of a connection between \wini\ and \lin, and the TCP four tuple that describes the connection: IP addresses and ports. Furthermore, the adversary must (2) obtain a valid sequence number (used by the spoofed source). The 32-bit sequence number, that is initialized to a random value, indicates `where' the data fits within the TCP connection stream. The receiver keeps a congestion window (\cwnd) for every connection, and discards a packet if its sequence number indicates that the data is outside \cwnd; see details in \cite{rfc793}. 

%\subsection{Circumventing Same Origin Policy and XSS}
%We show how a spoofing off-path attacker can inject malicious script so that browser runs it in the context of victim site, i.e., circumventing 

To circumvent the {\em same origin policy} \cite{rfc6454,Zalewski:web:sec:book}, the off-path attacker sends  forged responses for requests that \wini\ sends to another server \lin. This attack is facilitated in two phases: first the puppet opens a connection to the victim server, allowing TCP injection into this connection; then the puppet requests an object, allowing the attacker to send the script in a (spoofed) response.

In particular, this allows powerful {\em cross site scripting (XSS)}. XSS is one of the most critical attacks on web security, however, current XSS techniques depend on implementation vulnerabilities, usually of the site (e.g., \cite{XSS,Zalewski:web:sec:book}), and sometimes of the browser \cite{klein2005dom}. 
In contrast to typical XSS attacks, our attack does not rely on a server side or browser vulnerability. Moreover, since \mal\ can choose the server \lin, any persistent HTTP connection between \wini\ and a server is vulnerable (see above). Connections with HTTPS servers are not vulnerable to XSS since \mal\ cannot inject content into the cryptographically sealed session.

Furthermore, the XSS ability allows injection of requests, i.e., {\em cross-site request forgery (CSRF)} \cite{CSRF}.  This circumvents  existing defenses, such as the use of cookies, referrer header and hidden field, since all of these are available to the (injected) script (see \cite{CSRFPrevention}). The CSRF attack can be prevented by challenge response methods that require user involvement, such as password authentication or CAPTCHAs.

The XSS ability also allows  advanced phishing attacks. In particular, this provides efficient means for detection of browsing history, more effectively than previous techniques, e.g., \cite{conf/esorics/Kreitz11,conf/sp/WeinbergCJJ11}. 

%The `same origin policy' (SOP) defenses are a key to browser security, and date back to Netscape Navigator 2.0 \cite{SOPNetscape}. The same origin policy principle allows only very limited ways for sites in different domains to communicate. We show how \mal\ can perform XSS attacks and bypass the same origin policy. Furthermore, by facilitating XSS \mal\ can perform CSRF, successfully bypassing most defenses (see \cite{CSRFPrevention}); these include the (common) use of server provided hidden tokens which are attached to every request \wini\ sends to \lin. The reason is that by bypassing browser SOP, \mal\ gains access to data that \wini\ receives from \lin; hence, \mal\ is able to read this token.

%Since our attacks run from a browser script, we assume that \lin\ is an HTTP server. 

%Notice this attack fails against cryptographically-protected connections (SSL/TLS).  

% Furthermore, we show how \mal\ can perform cross-site request forgery (CSRF) \cite{CSRF} 

\subsection{Devastating DoS attacks}
An off-path attacker can use the knowledge of TCP parameters (IPs, ports and sequence numbers) in several ways, to attack the {\em availability} of a communication service. 

In 2004, Watson observed that BGP connections used a constant client port, and typically have very large windows; this makes it feasible for an off-path adversary to reset BGP connections \cite{watson2004slipping} (despite random initial sequence numbers). However, appropriate countermeasures make this attack inapplicable today \cite{rfc4987}. 

We show how a spoofing, off-path attacker, who controls a limited number of (weak) puppets, can deploy formidable DoS attacks, which so far were known to require  stronger attacker capabilities: the {\em Ack-Storm} attack \cite{AH11:AckStorm} and the {\em Optimistic-Ack} attack \cite{SBB05:OptAck}. Both are DDoS attacks, which use TCP control plane to generate excess amount of traffic. The Ack-Storm attack \cite{AH11:AckStorm}, is usually performed by MitM adversaries, possibly with limited eavesdropping abilities.  The Optimistic-Ack attack \cite{SBB05:OptAck} typically requires client cooperation (zombie) and persuades the server to send data in a high capacity, more than that allowed by \wini's link. 
%In the Ack-Storm attack, a MitM adversary injects two packets into the connection between \wini\ and \lin. After the injection, each packet that one peer sends to the other generates an Ack response back to the sender. This essentially causes a never ending sequence of acknowledgments. By injecting more than two packets the adversary initiates additional `Ack ping-pongs'. We show how an off-path adversary can perform an equally devastating Ack-Strom attack. 
%In the Optimistic-Ack attack the client (\wini) is a zombie controlled by the attacker; the attack changes \wini's TCP implementation to send acknowledgments for data that \lin\ had sent, but did not necessarily reach \wini. This convinces \lin\ to send data in high volumes, believing that his packets do reach \wini. By injecting Ack packets to the TCP connection we allow an off-path adversary to perform the Optimistic-Ack attack even without control \wini's TCP/IP stack.
Since in both these attacks, \mal\ injects data only to the TCP layer (and not to the application), these attacks also work when the victim servers use SSL.

Launching these attacks simultaneously on multiple client-server pairs may allow \mal\ to conduct an improved variant of the {\em Coremelt attack} \cite{conf/esorics/StuderP09} and congest a core link of the Internet, using only puppets.

%In the original Coremelt attack \cite{conf/esorics/StuderP09}, the adversary generates traffic between pairs of zombies in a bot-net under his control. A good choice of such pairs allows the attacker to load targeted core Internet links. We show how this attack can be improved: first, instead of zombies (fully controlled machines), we require that users enter a malicious website. Second, the adversary can choose the server, i.e., the peer for each of the users (clients). Instead of choosing from the bot-net itself, which is geographically centered in some cases, the adversary chooses web servers that also often have large bandwidths. The generation of traffic between the client and the selected server is by employing one of the two DoS attacks above. The vast deployment of Windows on client machines \cite{Wikimedia} makes these attacks practical. 
 
\subsection{Organization}
The next two sections focus on the application of the TCP injection technique. In Section 2 we present our off-path attacks on the confidentiality and integrity (authentication) of the communication between client and server, including the XSS, CSRF and phishing attacks. In Section 3, we present our off-path DDoS attacks. 

Sections 4-6 present the TCP injection technique itself. Section 4 presents the first step, which is exposing the server's sequence number. Section 5 continues the attack, to expose the client's sequence number as well. Section 6 discusses deployment challenges, improvements to meet these challenges, and experimental results. 

Finally, Section 7 proposes defenses against the attacks, and Section 8 presents a concluding discussion. 

%% Shor overview, still disorganaized.

%Our TCP injection attack assumes knowledge of a server $S$ IP and port, and injects data to a connection of $S$ with a client $C$ (in both direction, i.e., to $C$ and $S$).

%The once predictable initial sequence number, allowed blind adversaries (e.g., \mal) to setup `ghost' connections, i.e., TCP connections form spoofed addresses, the attack was first noticed in the 1980s, but gained attention only in 1995 \cite{morris1,Tsutomu}.

%\input{Model}
%\input{OnSequences}
\section{Off-Path Data Integrity Attacks}  \label{exploiting:S}

In this and the following section we present and empirically evaluate exploits of TCP injections. We focus on {\em long-lived-connection injection attacks}, where an off-path attacker learns the sequence numbers of an existing, long-lived TCP connection, between a given TCP client and server (identified by their IP addresses and ports). Motivated by these exploits, in Sections \ref{Injection:SeqExposure}-\ref{Injection:Experiments} we present and evaluate the technique we employ to study the sequence numbers. 

\ignore{
Indeed, even following reports by us and others (e.g., \cite{lkm:phrack:07}) identifying the vulnerabilities allowing TCP injections, these were not fixed. Vendors and developers seem to think that there are no significant threats and exploits due to injection attacks. We believe that this is wrong, and these vulnerabilities must be fixed as soon as possible, since they can be abused and exploited in different ways, causing significant damages. 
}

Specifically, we show critical exploits of long-lived connection injections. In this section we focus on two exploits: the first allows an off-path attacker to run a malicious script in the context of an arbitrary website of the attacker's choice, without depending on a vulnerability of the server (e.g., bug in input sanitization) or of the browser; this is a new, devastating type of XSS attack \cite{XSS,klein2005dom}. The second exploit allows the same attacker to present spoofed web-pages for clients. In the following section we show how off-path attackers can use long-lived connection injections to cause devastating DoS attacks. All exploits work in the same setting, %presented in Subsection \ref{Into:Contributions} and
illustrated in Figure \ref{fig:ourmodel}.

\subsection{Identifying the Victim Connection} \label{exploiting:S:conn}
To launch the long-lived-connection injection attacks, the attacker must identify a connection between the client and server which is defined by the IP addresses and ports of the participating peers.

The exploits use the puppet running on the client to {\em open} such (long-lived) connections. The server's IP and port are, of course, known. To find the client's IP and port, the puppet opens another TCP connection to the attacker and over it, sends packets to the attacker's machine. These packets contain the client's IP address. 

The final challenge is to detect the client port. Many clients, in particular, those running one of the Windows family operating systems, assign ports to connections {\em incrementally}. We use the puppet to open a connection to the attacker's remote site before and after opening the connection to the victim server, incremental port assignment allows the attacker to learn the client's port; see step A in the attack process described in Section \ref{exploiting:S:process} below. Client port exposing may fail if the puppet communicates with the server or attacker via a NAT device that randomizes the client port (since `external' ports are not incremental). In an ongoing research we investigate this problem and provide, details of a different technique that allows an off-path attacker to detect the client port in a connection. 

It remains to describe, in the following subsections, the unique aspects to each of the exploits and evaluate their impact.

\subsection{Off-Path Injection XSS (or: XSS of the Fourth Kind)}
In a {\em Cross-site scripting (XSS)} attack, the attacker causes the browser to run malicious, attacker-provided script (or other sandboxed code), with the permissions of scripts within a victim server web-page. 

In  \cite{klein2005dom}, Klein identifies three kinds of XSS attacks. In {\em persistent/stored} XSS attack, the script is received from the victim server, as part of the contents of a page stored by the victim server. In {\em reflection} XSS attack, the script is `reflected' by the victim server to the client, after the server receives the script from the browser (typically visiting a malicious website). Finally, in a {\em DOM-based} XSS attack, the script is received by the browser directly from the attacking server; a browser vulnerability (bug) causes the browser to consider the script as coming from some other victim server.

Both reflection and persistent/stored XSS attacks, exploit `bugs' in the web application. Well designed sites, using appropriate defenses such as Web Application Firewalls (WAF), should eliminate these attacks; see, e.g., \cite{XSS}. DOM-based XSS attacks do not require any bug in the site, but depend on bugs in the browser; the relevant known bugs were quickly patched by browser vendors.

Long-lived-connection injection attacks, allow a new, fourth kind of XSS attacks: {\em off-path injection XSS} attacks. In these attacks, the malicious script is sent by the attacker to the browser, with (spoofed) source IP address of the victim server. If the script it injected correctly, with correct TCP/IP parameters and within correct HTTP context, then the browser executes it in the context of the victim site. 

\vspace{-5pt}
\subsubsection{Attack Process} \label{exploiting:S:process}
\vspace{-5pt}
We next explain the technique we employ to use long-lived injection attacks to perform off-path XSS injections. Like our other exploits, we assume that the user visits a website controlled by the attacker from where he receives and executes a {\em puppet} (malicious script) \cite{AATA08:puppetnets}. The attack has five steps and proceeds as follows: 

\vspace{-5pt}
\subsubsection*{A. Form Connection, Expose Client Port} \label{exploiting:S:process:A}
\vspace{-5pt}
Puppet opens a new connection to a server controlled by the attacker, then a connection to the victim web server and finally another new connection to a server controlled by the attacker. Let the client port numbers that the attacker observes for the first and third connections be $p_1$, $p_3$. We use the counter property of Windows port assignment: if $p_3 = p_1+2$, then we assume that the client used port $p_1+1$ for the (middle) connection to the victim server. Otherwise, repeat. %Appendix \ref{appendix:clientport} provides a general technique that does not assume sequential port assignment for exposing the client port.

\vspace{-5pt}
\subsubsection*{B. Expose Connection Sequence Numbers} \label{exploiting:S:process:B}
\vspace{-5pt}
Puppet maintains the connection with the victim server alive by sending periodic requests for small objects. During this time, attacker runs the sequence exposure attack described in Sections \ref{Injection:SeqExposure}, \ref{Injection:AckEsposure}. If sequence exposing fails, restart entire attack. 

\vspace{-5pt}
\subsubsection*{C. Send `Dummy' Request} \label{exploiting:S:process:C}
\vspace{-5pt}
Puppet sends the victim server a request for some web page (over the same persistent connection), e.g., using an iframe, and informs the attacker on that request. Note that the puppet runs in the context of the attacker site; hence, the attacker and puppet can communicate and coordinate the attack without restrictions.

\vspace{-5pt}
\subsubsection*{D. Send Spoofed Response} \label{exploiting:S:process:D}
\vspace{-5pt}
Attacker sends spoofed response to the client, containing exact expected TCP parameters, and a web page containing the malicious script. 

\vspace{-5pt}
\subsubsection*{E. Script Execution} \label{exploiting:S:process:E}
\begin{figure}
	\vspace{-10pt}
  \begin{center}
    \includegraphics[width=0.45\textwidth]{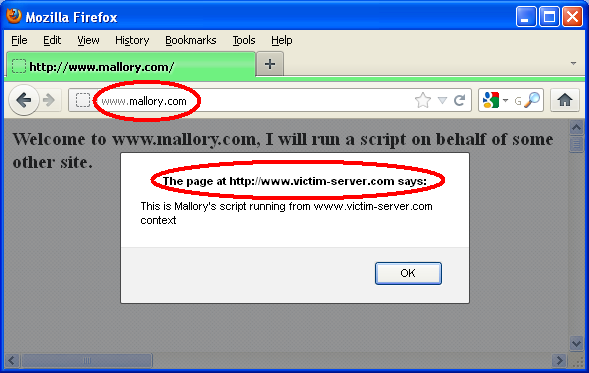}
  \end{center}
    \vspace{-10pt}
  \caption{An XSS Attack. \mal\ runs a script in context of \victimserv\ within Mozilla Firefox sandbox. The address bar indicates the user is at \malcom, but the message box context indication shows that the script (that \mal\ provided) runs from \victimserv.}
  \label{fig:XSSRun}
\end{figure}

\vspace{-5pt}
Browser receives the spoofed response as if it was sent by victim server, hence, executes script with permissions of the victim server. Figure \ref{fig:XSSRun} shows a successful run of this attack on the Mozilla Firefox browser.

\subsubsection{CSRF Exploit}

As indicated in \cite{XSS,CSRF}, once attackers succeed in an XSS attack, i.e., run a malicious script in the browser, in the context of a victim site, they can exploit it in many ways. In particular, such XSS attack allows attackers to send a forged (fake) request to the server on the user's behalf, i.e., a {\em cross site request forgery (CSRF)} attack, circumventing all known defenses against CSRF attacks for non-secured connections, except for (few) defenses requiring extra user efforts for submission of each (sensitive) request; see \cite{CSRFPrevention}. 

Note that since the attackers (cross site) scrips can read the entire response that the user receives from the victim web-server, they would even be able to circumvent advanced proposed defenses, which require new browser mechanisms. In particular, they can foil the {\em origin} header proposed by Barth et al. against CSRF attacks \cite{conf/ccs/BarthJM08}, as well as policy-based defense mechanisms against XSS, e.g., {\em Content Security Policy (CSP)} \cite{conf/www/JimSH07,conf/www/StammSM10}.

\subsection{Experimental validation} 

In this subsection we evaluate the applicability of the XSS attack on web-users. The client machine in the following experiments is protected by Windows Firewall. 

The success of the XSS attack depends on successfully exposing of the sequence numbers used in the connection the client has with the victim server. The success rate of the exposure technique that we employ (presented in Sections \ref{Injection:SeqExposure}, \ref{Injection:AckEsposure}) depends on the rate of packets that the client machine sends. In Section \ref{Injection:Experiments} we present another set of experiments that specifically evaluates the injection technique for different environments. In the measurements below, \wini\ sends 32 packets per second on average.

\begin{figure}
	\vspace{-10pt}
  \begin{center}
    \includegraphics[width=0.45\textwidth]{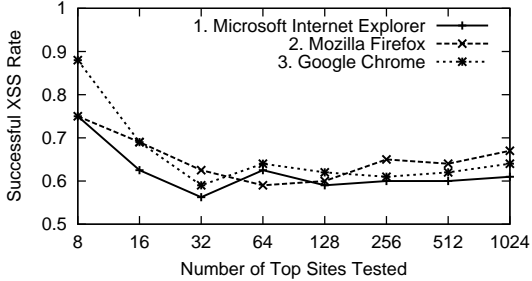}
  \end{center}
    \vspace{-15pt}
  \caption{Rate of successful XSS attacks on connections to popular sites.}
  \label{fig:XSSComparison}
\end{figure}

\begin{figure}
	\vspace{-10pt}
  \begin{center}
    \includegraphics[width=0.45\textwidth]{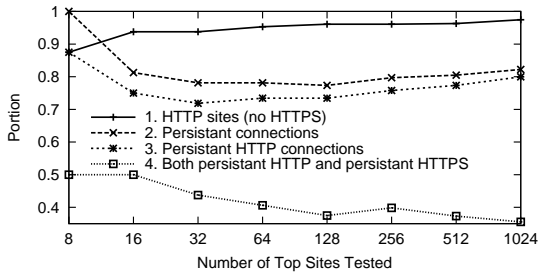}
  \end{center}
    \vspace{-15pt}
  \caption{The applicability of the injection attacks on various sites.}
  \label{fig:applicability}
\end{figure}

We tested whether connections with each of the top 1024 sites in Alexa ranking \cite{AlexaRanks} are vulnerable to off-path XSS attacks: our client machine connects to the attacker (\malcom), who then tries to run a script in context of one of the top sites. The script provides an indication of a successful injection by requesting an image from \malcom. Note that our attacker only communicates with the client, and does not have any interaction with the victim servers. In Figure \ref{fig:XSSComparison} we compare the results for three common browsers and observe that the attack is not browser-dependent. The immune connections are generally of the following types: (1) secured with SSL (HTTPS), this prevents the attacker from injecting his script to the connection (step D in the attack); (2) sites that do not use the HTTP keep alive option, this prevents the attacker from keeping the long connection with the server that is required to expose the sequence numbers (step B in the attack). In Figure \ref{fig:applicability} we provide distribution of the top 1024 sites in Alexa ranking; showing that 80\% percent of them appear vulnerable (line 3 in Figure \ref{fig:applicability}). A comparison of this result to those presented in Figure \ref{fig:XSSComparison} shows that the XSS attack was, in fact, successful on roughly 75\% of the sites that appear vulnerable. Among the vulnerable sites on which we ran a successful attack are {\em www.facebook.com}, {\em www.yahoo.com} and {\em www.amazon.com}.

\subsection{Web Spoofing/Phishing/Defacement} \label{exploiting:s:webspoofing}

In addition to the XSS exploits, attackers can use TCP injections to perform {\em web spoofing} (which is key to {\em phishing} attacks). Namely, the attacker
waits for the user to browse to some victim server, e.g., {\em http://www.bank.com}, and injects his data to the connection. In this attack, the attacker provides a spoofed version of the web-page to the client. This spoofing exploit can expose sensitive user-provided information such as passwords and may trick the user to download malware. An implicit assumption of this attack is that the initial web-page that the user receives, and which the attacker forges, i.e., {\em http://www.bank.com}, is not protected by SSL. This is the situation in most sites, which do not use SSL/TLS at all. 

The attack also works for many sites which do use SSL/TLS, but only via a link, e.g, to the login page {\em https://www.bank.com/login.php}. This approach is common since it reduces the load on the server by delaying setup of SSL connections until these are required (in the banking example, for login); see line 4 of Figure \ref{fig:applicability}\footnote{Line 4 of Figure \ref{fig:applicability} counts sites which have persistent connections, to both {\tt http} and {\tt https}. Note that some of these sites may not often use {\tt https}, while others may use {\tt https} but in a different domain.}. Web-spoofing can allow the attacker to circumvent the use of encrypted connections (SSL/TLS), using techniques/tools such as SSL-strip \cite{sslstripe}, i.e., replace links on the original page to phony pages (on the attacker's site). %Similarly, this can allow circumvention of advanced anti-CSRF defenses requiring user interaction and device-based user authentication mechanisms. 

To succeed in a web-spoofing attack, the attacker would best send the spoofed page as a response to a request made by the user (since then the page appears authentic to the user); hence, the attacker should be able to {\em detect} the request for the page and send a response. We solve this problem by having the puppet open a connection to the victim server in advance providing sufficient time to expose the sequence numbers used in the connection. We leave the connection open (by sending `dummy' requests periodically); and probe for user activity by identifying a change in the client's sequence number. In order to detect this change that indicates that the client had sent a request to the server, the attacker periodically conducts a {\em client-seq-test}; this test is a building block of the sequence numbers exposing technique and we provide its details in Section \ref{Injection:AckEsposure}. Briefly, the test allows the attacker to test whether the client sequence number is above some value; testing using the exposed (i.e., last known) value of client sequence number allows to detect such change. Once we detect such activity over the connection, we assume that the user had sent a request to the server for the home page and send a spoofed (modified) page.

This web spoofing technique assumes that the user opens the page for the victim-server while the puppet is still running, e.g., in a different tab of the same browser or in a zero-size iframe. Furthermore, it assumes that the browser employs connection sharing between different tabs, i.e., one TCP connection is used to communicate with the same server via several tabs of the browser. TCP connection sharing is employed by the current versions of Internet Explorer, Firefox and Chrome (and possibly other browsers).

Another assumption is that the user receives the attacker's response before the server's; this appears as a race that would be difficult to win for an attacker far from the client machine. However, the attacker can avoid this race by injecting data to the client (as the server) in advance: the injected data artificially increments the sequence number that the client expects from the server while the true server would still use the `normal' sequence number, causing the client to reject all data sent by the server.

\subsubsection{Example: Spoofing J.P. Morgan}

\begin{figure}
	\vspace{-10pt}
  \begin{center}
    \includegraphics[width=0.5\textwidth]{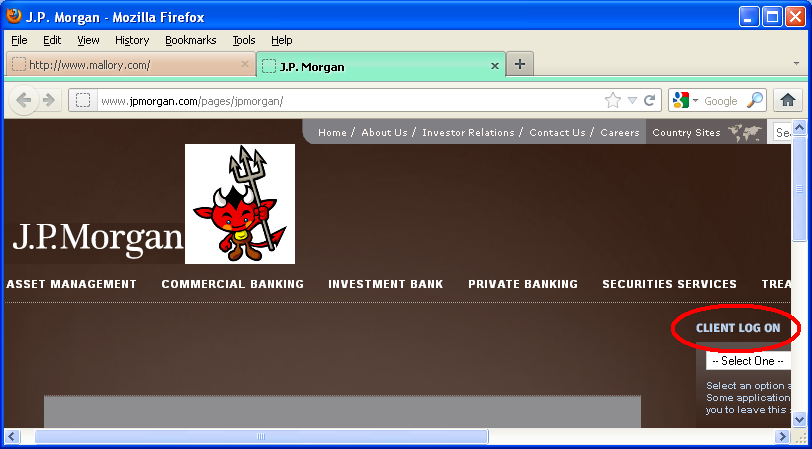}
  \end{center}
    \vspace{-15pt}
  \caption{Web Spoofing/Defacement Attack. \mal\ waits for the user to enter J.P. Morgan bank website, when he enters he injects a phony page. In this figure \mal\ added a devil image.}
  \label{fig:web-spoofing}
\end{figure}

The J.P. Morgan bank website is an example of a sensitive site that is vulnerable to this spoofing/phishing attack; it uses HTTP keep alive option and its homepage is not protected by SSL. Hence, this website is vulnerable to the web spoofing attack above. Figure \ref{fig:web-spoofing} shows the result of a successful web spoofing attempt: here the client has two tabs open in his browser. The current tab (in focus) shows the J.P. Morgan homepage that \mal\ provided; the devil image (that does not exist in the original page) indicates that this page is spoofed. J.P. Morgan homepage contains a client log-on link that in the original site switches to SSL. In the spoofed version, the link is to a web-page in \mal's site. In the other tab, the victim is in \malcom; this allows \mal\ to monitor the requests that the user (may) send J.P. Morgan and identify the correct time to inject the spoofed page.  

\section{Off-Path Denial-of-Service by Puppets} \label{exploits:DoS}
We next describe possible exploits of long-lived-connection injection attacks to disrupt network communication. Namely, we show how attackers can build on these attacks, to launch devastating Distributed Denial-of-Service (DDoS) attacks. As before, we consider an off-path IP-spoofing attacker, who also controls a significant number of `puppets', i.e., scripts running in browsers of unsuspecting users. As argued in \cite{AATA08:puppetnets}, it is relatively easy for attackers to control a large number of puppets in this way.  

However, puppets have limited ability to launch DDoS attacks. One reason are limitations placed by the browsers on the number of concurrent connections opened by the same web-page. Another reason is that, due to their execution within a sandbox, puppets can only use standard TCP connections. In particular, if the attack succeeds in causing congestion and loss, then TCP connections, including those of the puppets, will significantly reduce their window size (and hence their traffic rates). Therefore, while puppets can be used for DDoS attacks, their impact is much less than that of `regular' malware (zombies/bots). 

%In each of the following subsections we consider a different DDoS attack and show how by injecting packets into the connections between puppets and victim servers, the attacker can cause much more aggressive and effective attacks. 

In contrast to the data integrity attacks in the previous section, the attacks below use TCP {\em control plane} to cause congestion. Since there is no attempt to inject data to the application layer, even SSL/TLS protected sites are vulnerable. Hence, these attacks are applicable to websites that support persistent HTTP connections, with or without SSL/TLS, e.g., to about 80\% of the 1000 most popular sites (see line 2 in Figure \ref{fig:applicability}).

\subsection{Off-Path Optimistic Ack Attack} 
We first describe {\em Off-path Optimistic Ack}; this is a variant of the Optimistic Ack DoS attack  \cite{SBB05:OptAck}. These attacks cheat TCP's congestion control mechanism, causing senders to believe that most of the data they sent was already received, and hence that they can send more data (congestion window is not full), and also to increase the size of the congestion window. This can result in huge amplification factors, see  \cite{SBB05:OptAck}, unless servers use per-connection bandwidth quotas or use other defenses. % that in practice, the attack will be mostly limited by limitations imposed by the server on the throughput of a single connection. 
 
In both Optimistic Ack attacks (ours and the original \cite{SBB05:OptAck}), the attacker sends to the server acknowledgment packets (Acks), as if the client received all packets sent by the server (although packets are still in transit or even lost). As a result, the server continues sending information, with increasing window sizes (and hence rates). 

In the original attack \cite{SBB05:OptAck}, the client must run malware, with ability to send `raw IP' packets (i.e., not according to the TCP specifications). This is a significant requirement; recent operating systems make it harder for malware to obtain such ability. Furthermore, the original attack may be blocked by a firewall on the client side, by detecting the unusual high rate. Note that since by requiring only puppets, attacker is more likely to control enough clients to succeed in the attack, in spite of countermeasures such as per-connection quotas. %In addition, many servers may limit the impact of the attack by limiting its sending rate for each connection. Such limitation also applies to our variant, however, since we only requires a puppet, the number of attacking connections can easily be too large for this measure to be effective defense. 

Our long-lived-connection injection attack,  allows an off-path attacker to perform an off-path variant of the Optimistic Ack attack, as follows. The attacker only needs a puppet on the client machine to open the TCP connection with the victim server and learn the client port and sequence numbers. Following this, the puppet requests some large object from the server and is no longer needed; the attacker sends Ack packets as done by the client in the original attack, and - if not using SSL/TLS - the attacker can even send new request(s) if needed. Note that even if the client's firewall (detects the attack or for some other reason) begins blocking packets on this connection from both directions, this does not help since we provide the Acks to the server from the off-path attacker. Furthermore, RFC-compliant RST packets that the firewall (or client) may send, would be out of the server's window and hence ignored, and would not tear the connection. Server-induced verifications, by intentionally dropping or reordering packets periodically, as suggested in \cite{SBB05:OptAck}, seem one of the best (or only) defenses.

\subsection{Off-Path Ack-Storm DoS Attack}
In the original Ack-Storm DoS attack \cite{AH11:AckStorm}, the attacker needs to have some (limited) ability to eavesdrop on packets. From these packets, the attacker learns the TCP parameters (IP addresses, ports, and sequence numbers). Using these, the attacker sends two spoofed data packets, one to each of the two ends of the connection. 
According to the TCP specification, and in most TCP implementations, upon receiving an Ack for data that was not yet sent, TCP sends back a `duplicate Ack' - i.e., resends the previously sent Ack. As a result of receiving the pair of spoofed data packets, one at each peer, both peers begin sending acknowledgment packets to each other. Since these Acks acknowledge data which was actually sent by the attacker, not by the peer, then each of these Acks will only result in another duplicate Ack returned, and this process will continue indefinitely. %TCP specifies that the response to such packet is an Ack with the current sequence numbers, which causes the two parties to enter into infinite exchange of such Ack packets. 

The attacker can send additional data packets to the peers, causing additional ping-pong exchanges, quickly filling-up the channel capacity. This causes increased load on the networks; the fact that the packets involved in the attack are very short (just Acks),  makes the load on routers and switches even higher. Eventually, this causes packet losses, and legitimate TCP connections sharing the same links significantly reduce their rate. %; the attacker can easily send few more attack packets to compensate for any losses. %For details, see \cite{AH11:AckStorm}. 

The off-path Ack-Storm DoS Attack works exactly like the regular Ack-Storm DoS Attack, except for using the long-lived connection injection technique to allow the off-path attacker to learn the TCP parameters. Hence the attacker can run this attack, without requiring the ability to eavesdrop. %Notice that this attack involves only TCP control mechanisms, and does not involve any data; hence, it works equally well for SSL/TLS connections. 

\subsection{Off-path Coremelt Attack}

Since the two DoS attacks described above can be launched using only puppets, it follows that even relatively weak attackers may be able to cause large amounts of traffic from many clients spread around the network. This can cause high load on servers, routers and links. 

In particular, by choosing well the pairs of clients and servers between which the attacks are launched, the attackers can cause huge amounts of traffic to flow over specific `victim' backbone routers and links. These backbone networks, connecting large core ISPs (autonomous systems), have very high capacities; by sending enough traffic to a particular destination, attacker can cause queuing and losses in the connecting router. As a result, Internet connectivity may break - first for TCP connections and then even for UDP applications.  
We use the term {\em Off-path Coremelt Attack} for the resulting attack on core Internet connectivity, since it is an off-path variant of the Coremelt attack \cite{conf/esorics/StuderP09}. The Coremelt attack uses a large {\em botnet}, sending large amounts of traffic between pairs of the bots, with the pairs chosen intentionally so that huge amounts of traffic will flow over specific victim link/router. As shown by simulations in \cite{conf/esorics/StuderP09}, this can result in congesting the victim link/router, and even in breaking connectivity in the Network. 

In the Off-path Coremelt attack, the attacker will also congest a core link; however, instead of depending on pairs of zombies (bots), here the attack just requires a puppet at one end. The other end of the connection is a legitimate server, and to cause huge amounts of traffic, the attacker uses one of the two off-path DoS attacks described above. 

This attack has three advantages compared to the original Coremelt attack: (1) controlling a sufficiently large and correctly-dispersed set of puppets is easier than controlling a comparable set of zombies; (2) we only need to control one end of each connection (the puppet), not both ends; and (3) since we use adversary-chosen servers, these can have very high bandwidth, higher than available to most bots.  

\subsection{Experimental Evaluation}
\begin{figure} [h]
  \begin{center}
    \includegraphics[width=0.35\textwidth]{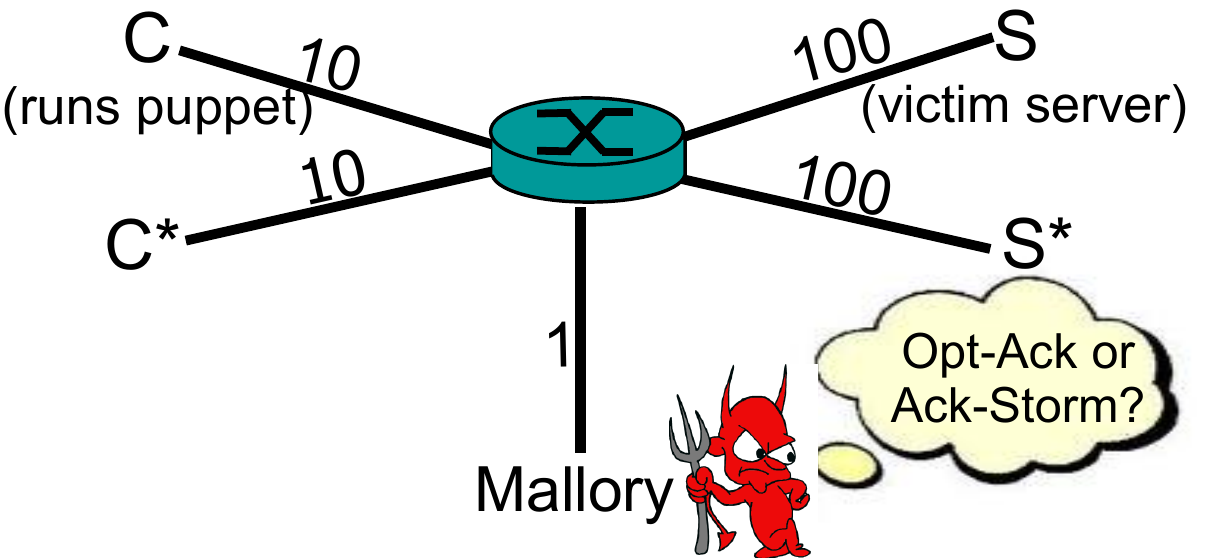}
  \end{center}
  \vspace{-15pt}
  \caption{Network environment for testing the DoS attacks. Each link specifies its capacity in mbps.}
  \label{fig:dosmodel}
\end{figure}

We used the topology illustrated in Figure \ref{fig:dosmodel} to test the off-path denial of service attacks we presented. In our tests we assume that \mal\ runs a puppet on \wini\ and can inject data to the TCP connection between \wini\ and \lin\ (a connection that \mal\ caused \wini\ to establish).

We evaluate the attacks by measuring the degradation of service that they cause to other legitimate connections. We consider different round-trip times (RTTs) for the legitimate connections we measure: the longer the RTT is, the greater the congestion windows are. Since every loss halves the congestion window, a more significant effect is observed when RTT is high. Furthermore, a higher RTT implies that it will take more time for the sender to detect a packet loss and retransmit. The base line to which we compare the effect of these attacks is line 1 in Figure \ref{fig:AckStromOptAck} which illustrates the time it takes \wini* to receive a 50MB file from \lin* under normal conditions.

\vspace{-15pt}

\subsubsection{Off-Path Optimistic Ack Evaluation}
\begin{figure}
	\vspace{-10pt}
  \begin{center}
    \includegraphics[width=0.45\textwidth]{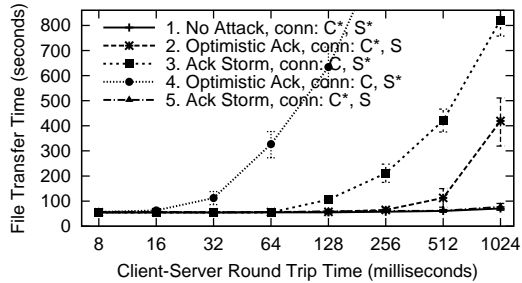}
  \end{center}
   \vspace{-20pt}
  \caption{Evaluation of Ack-Strom and Opt-Ack DoS attacks. The legend indicates for each line the type of attack and the peers in the legitimate connection 
  (that \mal\ aims to degrade). Measurements are the average of 50 runs, error-bars mark the standard deviations.}
  \label{fig:AckStromOptAck}
\end{figure}

In this attack we aimed to clog the \lin's link: \mal\ uses \wini\ to request some large file from \lin\ and then performs the Optimistic Ack attack, persuading \lin\ to send data to \wini\ at high a rate. We evaluated the effect of this attack by measuring the degradation of service in a connection that \lin\ has with some other client \wini* who tries to download a 50MB file. Lines 1 and 2 in Figure \ref{fig:AckStromOptAck} illustrate our results. Notice the significant difference in attacker and server link capacities; the amplification ratio measured in this attack is $78$ (for every byte that \mal\ sends, \lin\ sends approximately 78 bytes). Furthermore, the attack also clogs \wini's link, as shown by line 4.

\subsubsection{Off-Path Ack-Storm Evaluation}

We use the Ack-Storm attack to congest \wini's link and measure the effect on a different connection that he has with some other server \lin*. In order to congest the link, \mal\ creates a new Ack `ping-pong' every 100 ms; to create each ping-pong \mal\ sends only two short (40B) packets. In the connection with \lin*, \wini\ tries to download a 50MB file (similar to the previous experiment). Lines 1 and 3 compare the file transfer time at a normal time, to that when the Ack-Storm attack takes place. This attack requires much less effort and lower bandwidth than the Opt-Ack attack, i.e., has much higher amplification ratio; but is limited by the client bandwidth (which is typically lower than the server's) this limitation is illustrated by line 5 which is very similar to line 1 (normal conditions). 

\section{Server Sequence Number Exposure}\label{Injection:SeqExposure}

In this and the following section we describe the sequence exposure attack where a off-path adversary, \mal, learns the current sequence numbers of a TCP connection between \wini\ and \lin.

We present a two phase attack: first, in this section we describe how \mal\ learns the server's sequence number, \textit{sn}, which \lin\ will use in the next packet sent to \wini. In the second phase, presented in the following section, we show how given \lin's sequence number (\textit{sn}), \mal\ efficiently obtains the acknowledgment number that \wini\ expects; this acknowledgment number is the sequence number that \wini\ will next use in packets sent to \lin.
In both phases \mal\ communicates only with \wini\ and aids a side channel feedback that she receives from \wini.
%A similar technique was presented in the phrack magazine in 2007 \cite{}

The attacks that we describe in both sections assumes that \mal\ had previously identified \wini\ and \lin's IP addresses and ports; see details on how \mal\ exposes these parameters in Section \ref{exploiting:S:conn}.

\subsection {The Server-Sequence Test}
%\subsection{Obtaining \lin's Sequence Number} \label{Injection:SeqAck:Seq}
This subsection presents the {\em server-sequence test} that allows \mal\ to test whether some sequence number, \textit{sn}, is within the flow control window (\cwnd) that \wini\ keeps for packets he receives from \lin. The key observation is that when a connection is in the established state, the recipient's handling of an {\em empty acknowledgment} packet (i.e., acknowledgment with no additional data) differs in case that it is not within \cwnd. The difference depends on the 32-bit sequence number and allows \mal\ to search for a valid sequence number. The following paragraphs explain how:

Packets that specify an invalid sequence number (i.e., outside the recipient's \cwnd) cause the recipient to send a duplicate Ack (for the last valid packet the recipient received). However, if the sequence number is within \cwnd, then the receiver does not send any response; the reason is that replying with an Ack in this case would start a never-ending series of acknowledgments.

\begin{figure}[htb]
	\centering
		\includegraphics[width=0.48\textwidth]{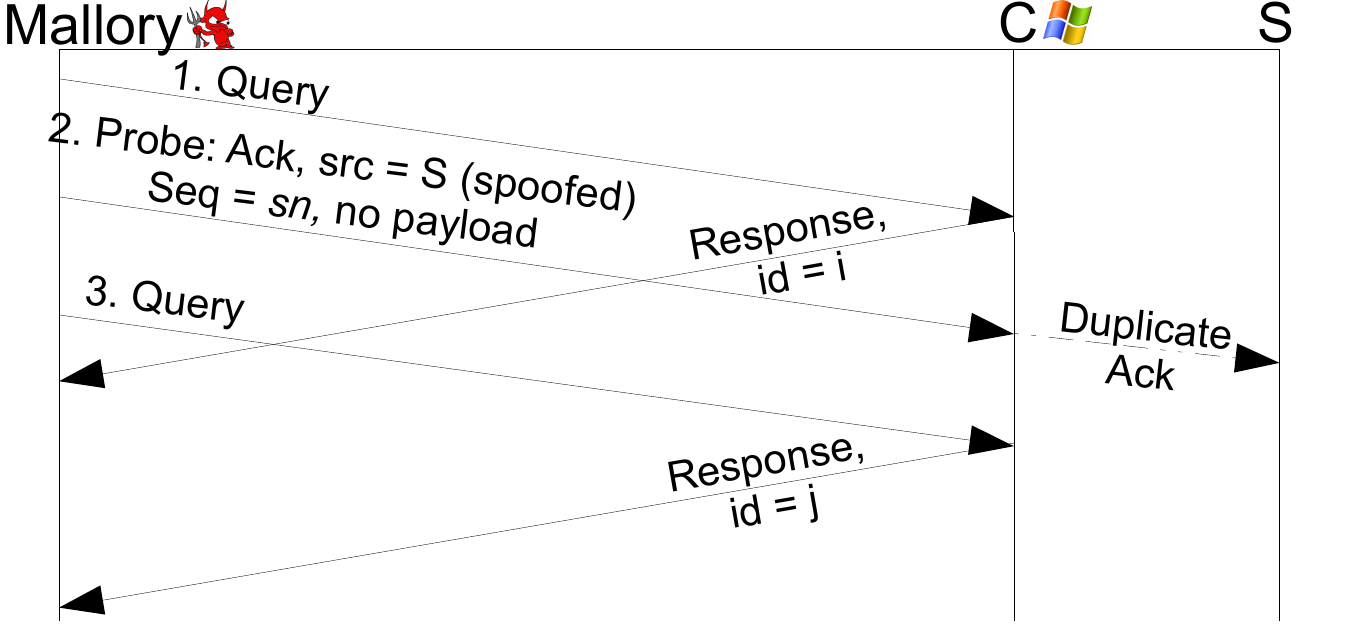}
	\caption{Server-Sequence Test. The dashed arrow marks the duplicate Ack that \wini\ sends \lin\ only in case that \textit{sn} is not within the flow control window. 
	If \textit{sn} is within \wini's \cwnd, then \wini\ does not send that packet and $j - i = 1$.}
	\label{fig:Seq}
\end{figure}

The {\em server-sequence test}, illustrated in Figure \ref{fig:Seq}, has three steps: in the first and third steps \mal\ sends a {\em query} to \wini; this is some packet that causes \wini\ to send a response packet back to \mal\ who then saves the IP-ID value in the response. In Section \ref{Injection:Queries} we show how \mal\ can use the legitimate TCP connection she has with \wini\ to implement queries and responses (since \wini\ is in \malcom). In the second step, \mal\ sends \wini\ a probe: this packet is spoofed and appears to belong to \wini's connection with \lin. The probe in this test is an empty Ack packet that leverages the observation above. 

When \mal\ receives the responses (for steps 1,3); she uses the IP-IDs they specify, $i$ and $j$, to learn $x = j - i$, the number of packets that \wini\ had sent between the two queries. Since the Windows IP-ID implementation is a single counter for all destinations (incremented on every packet that \wini\ sends), \mal\ learns that \textit{sn} is within \wini's \cwnd\ if $x = 1$, i.e., \wini\ did not send any packet between the two queries (see Figure \ref{fig:Seq}). 

\vspace{-15pt}

\subsection{Learning Process} \label{Injection:SeqExposure:learning}
The probability that a sequence number that \mal\ tests is within \wini's \cwnd\ is $\frac{\textit{wnd-size}}{2^{32}}$ (since \cwnd\ is $32$ bits long); this resembles to the TCP RST forgery attack \cite{watson2004slipping} where the attacker sends reset packets with arbitrary sequence numbers. The value of $\textit{wnd-size}$ is therefore important; Gont also points out the significance of the flow control window size in his security assessment for TCP \cite{Gont11TCP}.
%The greater the connection's round trip time and \lin's rate, the larger $\textit{wnd-size}$ is.

%If ${\textit{wnd-size}}$ is large, e.g., close to $2^{16}$ bytes, then this probability is about $2^{-16}$.

\mal\ conducts the server-sequence test until he identifies a sequence number within \wini's \cwnd. Each test is for a different sequence number which is of distance $\ests$ from the previously tested one, where $\ests$ is an estimation of \wini's $\textit{wnd-size}$. Namely, \mal\ tests sequence numbers $0, \ests, 2\ests$, etc. Generally, if $\ests < \textit{wnd-size}$, then \mal\ would test redundant sequence numbers, and if $\ests > \textit{wnd-size}$, then \mal\ may not find a valid sequence number. In our attacks (presented in Sections \ref{exploiting:S}, \ref{exploits:DoS}) we use the puppet to request some large resource (or few small resources) over the connection with \lin\ before initiating the sequence exposure attack; the response increases \wini's $\textit{wnd-size}$. We increase $\textit{wnd-size}$ until it is approximately $2^{16}$. 
Once a sequence number within \cwnd\ is detected, \mal\ conducts a binary search (over the possible \ests\ sequence numbers) to identify the exact beginning of \cwnd, which is the next server sequence number.

In Section \ref{Injection:Experiments} we provide an empirical evaluation of the complete sequence exposure technique.

\ignore{
\subsubsection{Numeric Example}
Assume that $\textit{wnd-size} = 2^{16}$ and $\ests = 0.75 \cdot 2^{16}$, then \mal\ would cover the entire sequence number space after $\frac{2^{32}}{0.75 \cdot 2^{16}} \approx 2^{16.5}$ iterations. Note that the number of required iterations decreases linearly to the size of \wini's flow control window; thus, if $\textit{wnd-size}$ is small, e.g., $2^{12} = 4\text{KB}$ bytes, then \mal\ would need to test $2^{20.5}$ sequence numbers. Since every test only includes three short packets, \mal\ would need to send $14-60$ MB of data, depending on the size of \wini's flow control window.
}
\section{Client Sequence Number Exposure}\label{Injection:AckEsposure}

In recent Windows client versions (from XP SP2 and onwards) the recipient uses the acknowledgment number, that is specified within TCP packets, together with the sequence number to verify that a packet is valid. In order to inject a packet to the TCP stream, \mal\ must specify an Ack number that is within \wini's transmission window; i.e., Ack for new data that \wini\ had sent. The black area in Figure \ref{fig:AckMap} represents the `acceptable' acknowledgment numbers (transmission window). In this section we show how to take advantage of Ack number validation to expose the client's sequence number.

\subsection{The Client-Sequence Test}

Similarly to the test we presented in the previous section, we build a three step {\em client-sequence test} where the first and last steps provide \mal\ with the current value of \wini's IP-ID. In the second step \mal\ sends a spoofed probe, \wini's response to this probe depends on the Ack number \mal\ specifies.

\begin{figure}[htb]
	\centering
		\includegraphics[width=0.481\textwidth]{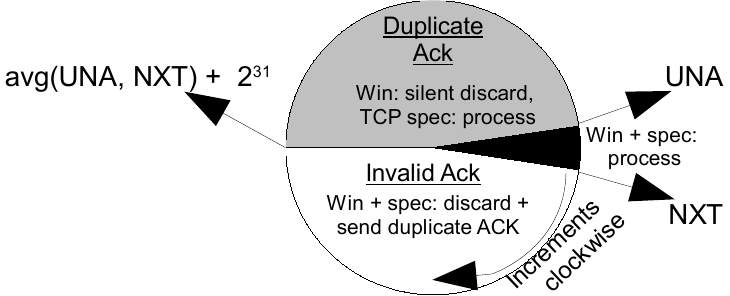}
	\caption{Ack Number Map. UNA is the lowest unacknowledged sequence number, NXT is the next sequence number that \wini\ will send. Acknowledgments for data in the {\em gray} area are considered duplicates, acknowledgments for data in the {\em white} area are for unsent data, i.e., invalid. In the {\em black} area are sequence numbers for sent un-Acked data. Note that the 32-bit Ack field is cyclic.}
	\label{fig:AckMap}
\end{figure}

The test is derived from another observation from the TCP specification \cite{rfc793} (Section 3.9, page 72). The relevant statement refers to an acknowledgment packet that carries data and contains a valid sequence number; i.e., success in the previous server sequence exposing phase is required to initiate this phase. The specification distinguishes between two cases regarding the acknowledgment number in the packet, see illustration in Figure \ref{fig:AckMap}. 

{\bf Case 1:} the packet contains a duplicate Ack (gray area in Figure \ref{fig:AckMap}), or acknowledges data that was sent, but not already acknowledged (black area in Figure \ref{fig:AckMap}). In this case the recipient is supposed to continue processing the packet regularly (see \cite{rfc793}). However, a Windows recipient (e.g., \wini) silently discards the packet if it is in the gray area (since acknowledgment is invalid); otherwise (black area) its data is copied to the received buffer for the application.

{\bf Case 2:} In the complementary case that the acknowledgment number is for data that was not yet sent (white area in Figure \ref{fig:AckMap}), the recipient discards the segment and immediately sends a duplicate Ack that specifies his current sequence number, NXT.

%These actions are summarized in Figure \ref{fig:AckMap}.

Hence, when \wini\ receives an acknowledgment packet that specifies an acceptable sequence number, i.e., within his flow control window (\cwnd), then:
(1) in case that the specified Ack number is after UNA, \wini\ sends an acknowledgment; either since data had arrived (black area), or since the packet acknowledges unsent data (white area). (2) In case that the Ack number is before UNA (gray area), then \wini\ (running Windows) discards it.

The probe which we use in the client sequence test specifies the acknowledgment number to be tested (\textit{an}) and has two important properties: (1) the probe packet specifies \textit{sn}, a sequence number that is within \wini's \cwnd\ (discovered in the server sequence exposing phase); (2) the probe packet specifies data. 

\begin{figure}[htb]
	\centering
		\includegraphics[width=0.48\textwidth]{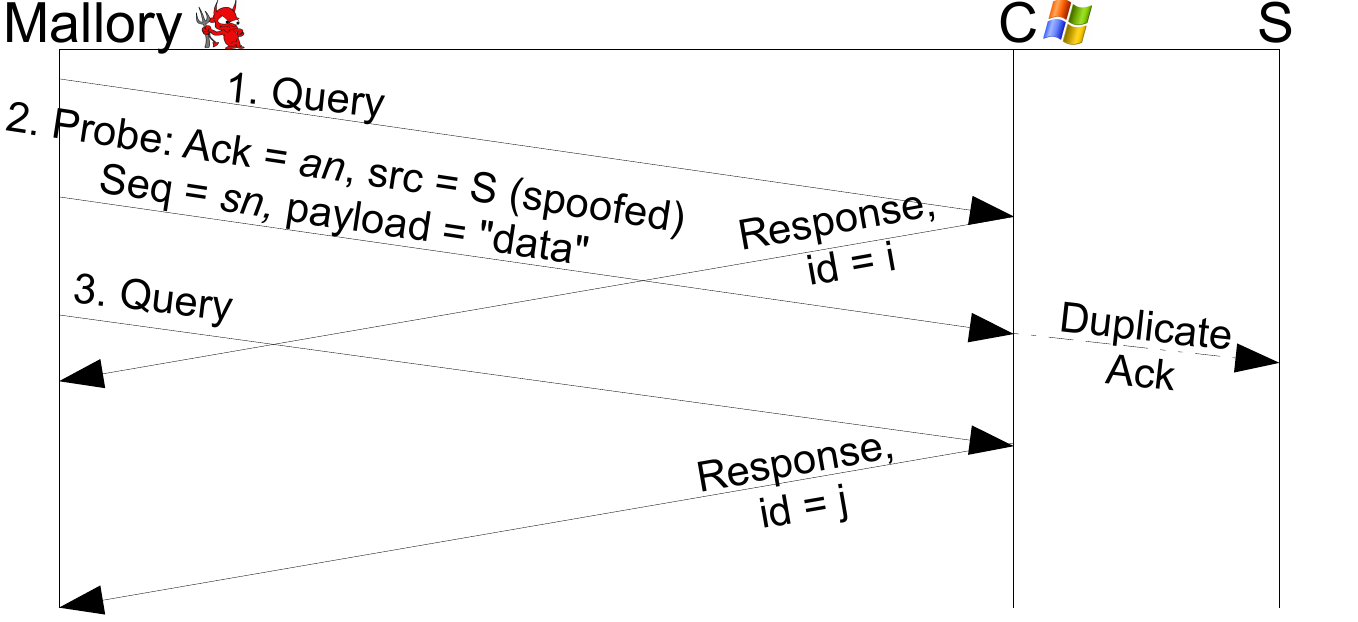}
	\caption{Client-Sequence Test. The dashed arrow marks the duplicate Ack that \wini\ sends \lin\ in case that \textit{an} is in the white area in Figure \ref{fig:AckMap}. If $j - i = 1$, then \wini\ did not send that packet; in this case \mal\ identifies that \textit{an} is below UNA (i.e., in the gray area in Figure \ref{fig:AckMap}).}
	\label{fig:AckTest}
\end{figure}

\subsection{Ack Number Binary Search} \label{Injection:SimpleBinary}

The client-sequence test allows \mal\ to conduct a binary search for the acknowledgment number that the client expects.
If the client-sequence test for the acknowledgment number $\textit{an}$ indicates that \wini\ did not send any packet between the two queries, then $\textit{an}$ is before UNA (in the gray area in Figure \ref{fig:AckMap}). Otherwise, \mal\ concludes that $\textit{an}$ is after UNA (in the black or white area in Figure \ref{fig:AckMap}). %In case that $\textit{an}$ number is before UNA and \wini\ sends an `independent' packet between handling the packets in the test, then \mal\ errors.

The gray and white areas in Figure \ref{fig:AckMap} are of equal size, and the black area (sent bytes without acknowledgment) is usually relatively small. This allows \mal\ to perform a binary search for UNA; each time eliminating approximately half the possible numbers. The 32-bit length of the Ack field (32 bits) implies that there are $32$ iterations.

\section{Sequence Exposure in Practice} \label{Injection:Experiments}

In this section we discuss the applicability of the sequence exposure technique in practice; we assume the model presented in Section \ref{intro}.

\subsection{Implementing Test Queries/Responses} \label{Injection:Queries}

The server and client sequence tests we described in Sections \ref{Injection:SeqExposure} and \ref{Injection:AckEsposure} use packets that \mal\ receives from \wini\ to learn the effect of the (spoofed) probe packet. \mal\ can persuade \wini\ to send her such packets by using the legitimate TCP connection she has with \wini: a query is some short data packet that \mal\ sends to \wini, the response is the \wini's acknowledgment sent back to \mal.

This method allows \mal\ to bypass typical firewall defenses since all packets in the test appear to belong to legitimate connections (requests to \wini-\mal\ connection, probe to \wini-\lin\ connection). Specifically, Windows Firewall does not filter this technique.

\subsection{Detecting Packet Loss} \label{Injection:packetlossdetect}

In order to succeed in sequence exposing, \mal\ must identify when test packets are lost since the corresponding test will yield a wrong result. For instance, if a probe is lost, then its test will indicate that \wini\ respond to the probe and mislead \mal. 

\mal\ detects a lost probe by repeating tests that indicate the client did not send a response (i.e., when $j-i = 1$). There should be only few such tests: one when probing for the server's sequence number and about sixteen during the binary search for the client sequence.

\mal\ detects lost queries and responses by employing TCP congestion control. Since we implement the queries as data sent on a TCP connection, we are able to detect a lost query similarly to TCP congestion control mechanism: if \mal\ receives several duplicate Acks, then she assumes that the corresponding query (a data packet) was lost. In this case \mal\ repeats all the tests that she performed between the corrupt test and until its detection. \mal\ detects a lost response by identifying that no Ack was received for one of the queries (instead an accumulative Ack was received); in this case she just repeats the invalid test.

\ignore{If such a packet is lost and the event is undetected, then future tests are corrupted: say the post probe query of test (0) was lost, consider the following two tests. The response for the pre-probe query of test (1) appears to \mal\ to belong to test (0); next the post probe query and pre-probe query of tests (1) and (2) will be sent sequentially, and (probably falsely) indicate that there was no response to the probe from test (1). This error would detect when \mal\ verifies this result as we presented in the previous paragraph. However, it will induce significant performance penalty if the probability for loss is high.}

\subsection{Test Errors} \label{Injection:errors}
The sequence exposure process uses the global IP-ID to determine whether a probe caused \wini to respond. However, since every packet that \wini\ sends increments the IP-ID, errors may occur. 

Such errors can appear only in tests where \wini\ does not respond to the probe: if \wini\ sends a packet in response to the probe, then the IP-ID is incremented, and the difference in IP-ID values of the responses that \mal\ sees is at least $2$; i.e., in this case \mal\ always concludes that \wini\ had sent a packet \footnote{We assume that \mal\ can send these packets with (small) inter-packet delay, such that reordering in the test packets is a rare.}. 

Hence, there are only few tests where an error is possible: during server sequence exposure only one test should indicate that no packet was sent, i.e., that \mal\ found a sequence within the recipient flow control window. \mal\ then conducts a binary search over the values in the flow control window to find the exact sequence number.
There are approximately 16 iterations to this binary search, on average, half of these indicate that \wini\ does not send a packet in response to the probe. Similarly, the binary search for the client-sequence includes 32 iterations, on average 16 tests should indicate the \wini\ did not send a packet. %In Appendix \ref{errorrate} we analyze the probability that \mal\ errors in one of these `sensitive' tests and show that it is low (e.g., if \wini\ sends on average $32$ packets per second, then the error probability is about 6\%).

\subsection{Experiments}

In this set of measurements we evaluate the sequence exposure technique; in Sections \ref{exploiting:S} and \ref{exploits:DoS} we evaluate the full attack (that requires sequence exposing and different successful `meaningful' injections). The server in these measurements is runs Apache, and the client is an up to date Windows machine (protected by Windows Firewall).

\begin{figure}[htb]
	\centering
		\includegraphics[width=0.42\textwidth]{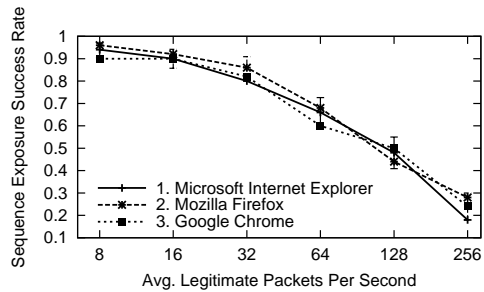}
	\caption{TCP sequence exposure success rate. Each measurement is the average of 50 runs, error bars mark standard deviations.}
	\label{fig:Attack}
\end{figure}

Figure \ref{fig:Attack} illustrates the success probability for different packets per second averages and when the puppet runs on different browsers. The average time for a successful sequence exposure is 102 seconds (standard deviation 18 seconds); this is the estimated time we require the client to stay in the attacker's site to conduct cross site scripting and initiate denial of service attacks (see Sections \ref{exploiting:S} and \ref{exploits:DoS}) \footnote{The web spoofing attack presented in Section \ref{exploiting:s:webspoofing} requires that the victim will stay in the attacker's site until he accesses the web-page that attacker wishes to spoof.}. Attacker and client bandwidths are respectively 1 and 10 mbps. 

In appendix \ref{appendix:comparison} we provide a detailed comparison between our sequence exposure technique to the one previously presented in \cite{lkm:phrack:07}.

\section{Defense Mechanisms} \label{Injection:Defenses}

The attacks in this paper relay on successful exposure of the sequence numbers, the technique we presented for this task uses the global counter property of the IP-ID implementation in Windows machines. Deployment of IPv6 mitigates the IP-ID attack vector since the IPv6 fragmentation header (that specifies the IP-ID) is only present in fragmented packets. In most implementations, TCP employs path MTU discovery to avoid IP-fragmentation. Hence, TCP connections over IPv6 are usually immune to our attacks. 

In this section we propose defenses that prevent off-path sequence exposing. Our mechanisms are of two types, those deployed at the client-end, and those at the server-end. Each mechanism blocks the attack even if the other side is unprotected; i.e., servers and clients can independently protect themselves. % without cooperation from the remote peer.

\subsection{Server-End Defense}
The defense at the server end uses feedback that the client machine (that runs the puppet) involuntarily sends as a side effect of the ID-exposing process. For every wrong guess of the server sequence number, the client sends the server a duplicate Ack (see section \ref{Injection:SeqExposure} and Figure \ref{fig:Seq}). We monitor the acknowledgments that the server receives to detect the attack. The following two firewall rules provide indications of an ongoing sequence exposure attempt.

The first rule verifies that in no point in time the server receives more Ack packets than the number of un-Acked data packets that he had sent to the client. However, the adversary can trick this rule by using the puppet to request some data from the server, the short window of un-Acked packets (`in transit') allows testing few sequence numbers. \ignore{Tricking the server in this way is difficult since the client also sends normal Acks when the server data arrives and as the attack continues, the number of Ack packets that the attacker causes the client to send increases; and so attacker must convince the server to send the client data in increasing rate (so that more packets will be in transit) to remain undetected \footnote{During sequence exposing, the attacker tests approximately $2^{15}$ sequence numbers}.}

We can further improve detection of the sequence exposure attack for connections that employ the TCP selective Ack option. This option is used by most modern browsers, including Internet explorer, Firefox and Chrome. A selective Ack specifies the sequence numbers of out-of-order data that was received by the sender and allows to distinguish between a duplicate Ack generated by a network loss and a duplicate Ack due to sequence exposure attempt. 

The second rule is enabled when the selective Ack option is used; it verifies that no two sequential Ack packets that the server receives are identical. Normally, a duplicate Ack is a result of a packet, $p$, that arrives out-of-order. In this typical case, $p$'s sequence number must still be in the recipient's flow control window; hence, $p$'s data is queued and the recipient sends a duplicate Ack to the source. The selective Ack attached to this feedback notifies that $p$'s data was received. However, in the case of a sequence exposure attack, most of the probes that the attacker sends are out of the recipient's (client's) flow control window and are discarded. Therefore, the duplicate Ack response to a probe is identical to the previous Ack that the server had received.

After several indications of an attack, the firewall tears down the connection. Note that abuse of this mechanism to cause the server to close a legitimate connection with one of his clients requires the adversary to send to the client a probe that specifies the correct connection four tuple. However, since in contrast to our attacks, the off-path attacker does not create the legitimate connection (that she tries to tear down), it is challenging to expose the connection parameters (addresses and ports).

\subsection{Client-End Defense}

In this subsection we propose modifying the IPv4 identifier at the client's firewall (to replace the global counter). Since the identifier is only used by the recipient to match packet fragments, when a packet arrives at the sender's firewall, it can modify the IP-ID field without any implications on the sender or recipient (even if the packet will be fragmented later on the route). When a packet arrives in fragments at the firewall, then it must map all fragments of the same packet (those that specify the same reassembly four tuple - IP addresses, protocol and IP-ID) to the same identifier.

The first, intuitively appealing direction seems to be using random identifiers. However in IPv4 this is not recommended, according to the birthday paradox, in roughly $1.2\sqrt{2^{16}} = \left\lceil 1.2\cdot2^8 \right\rceil = 307$ packets there will be a repetition (since the field is $16$ bits long), which would cause fragments of one packet to be mis-associated with others, and hence cripple performance.

The IP standards \cite{rfc791,rfc2460} specify that IP fragments are associated with a packet according to four parameters: source and destination addresses, transport layer protocol (e.g., TCP), and identifier. Therefore, a simple solution would be that each source, destination, protocol tuple will be associated with a different identifier counter, initialized by a keyed pseudo random function $f$, i.e., the initial identifier is $f_k(\textit{source, dest, protocol})$. In Linux, the choice of IP-ID is similar, but is only based on the source and destination addresses. % \footnote{Notice that keeping per-connection counters (i.e., per $<$source IP:port, dest IP:port, protocol$>$) may not work well: suppose the counters for two connections of same $<$source IP, dest IP, protocol$>$ would by chance reach the same range, they will frequently collide, which may result in packet losses.}. 

FreeBSD supports using random IPv4 IDs which are permuted locally: a packet is assigned with a random IP-ID that was not specified in one of the recent (8192) packets that were sent \footnote{The default FreeBSD configuration uses a globally incrementing IP-ID, as in Windows.}. Both Linux and FreeBSD approaches immune the TCP connection to our attacks. 

\section{Conclusions} \label{Injection:Conclusions}
In this work, we show that the folklore belief that TCP is secure against spoofing-only, off-path attackers is unfounded. We show practical, realistic  %significant attacks against privacy (traffic analysis), very feasible to deploy, as well as 
injection attacks. 
%, all in realistic settings (firewall, benign traffic, etc.). 
We further show that this allows crucial abuses, breaking the same-origin policy defense which is critical to web security, and allowing devastating DoS attacks.

%Our attacks exploit side-channels, so far mainly used in attacks against cryptographic systems, showing they can be very significant for network security attacks. We present several defenses against the specific attacks and side channels we found, see Appendix \ref{Injection:Defenses}; firewalls, TCP implementations and specifications should be modified. 

%Future work may further improve our attacks using these side channels (if not blocked), and possibly find more side channels and probes, using our `query-probe-query' attack pattern (or another attack). In particular, notice that our results already allow detection of traffic rates for Window systems, and it may be feasible to extend this to non-Windows systems. This may allow, e.g., trace back of TOR connections \cite{conf/uss/DingledineMS04}. 

One important conclusion is that Bellovin \cite{Bellovin:Look:Back:at:TCP:IP:Security} was right: TCP was never designed for security, and should not be expected to provide it. To ensure authentication and confidentiality, even against (only) spoofers, we should use secure protocols such as  SSL/TLS \cite{rfc2246} or IPsec \cite{rfc4301}. SSL/TLS may not suffice to prevent (lower-layer) attacks such as the DoS attacks presented in this paper (Section \ref{exploits:DoS}); to prevent these too, we should use lower-layer security mechanisms, preferably IPsec or other mechanisms, e.g., see \cite{rfc4953,conf/esorics/GiladH09}. 
%Notice that higher-layer security mechanisms such as SSL/TLS \cite{rfc2246} will prevent injection to TCP connection, but will not prevent the presented attacks on privacy and Denial of Service (e.g., RST forgery).  

A potentially more controversial conclusion is that basic vulnerabilities should be investigated and fixed, even before demonstration of a complete, practical, exploit. In this paper, we went into great length to prove the practicality of the vulnerability, since earlier results were considered as `impractical'. We believe that the network security community should adopt a more prudent approach, publishing and addressing issues and potential vulnerabilities, without waiting for a complete exploit. Compare the approach in the cryptographic community, where even yet-theoretical attacks are taken into account, published and motivate design of improved ciphers.

% ---- Bibliography ----
%\bibliographystyle{abbrv}
%\bibliographystyle{plain}
%\bibliographystyle{IEEEtran.bst}
%\bibliographystyle{spmpsci}
%\bibliographystyle{h-elsevier2.bst}
%\bibliography{rfc,LOT,spamplus,DoS,AmirCryptAndSec,dot,miscellaneous}
\bibliographystyle{plain}
\bibliography{NetSec,rfc,miscellaneous}

\appendix
\section{Comparing Performance of TCP Injection Attacks} \label{appendix:comparison}

In this appendix we compare the existing approach and technique for TCP injection presented in \cite{lkm:phrack:07} to those presented in this paper.
The significant difference between the two approaches is that \cite{lkm:phrack:07} injects data to a legitimate existing connection between two peers (\wini\ and \lin)
where in this paper we use a puppet to create the victim connection. This difference has three implications we describe below. 

First, the attacker must identify the connection between \wini\ and \lin\ and expose its parameters (IP addresses and ports). In \cite{lkm:phrack:07} attacker is assumed to have previous knowledge of the client and server addresses as well as the server's port; in this paper we assume only knowledge of the server's address and port which are usually available. In order to expose the client's port, in \cite{lkm:phrack:07} the attacker performs a variant of the idle scan, indirectly scanning all possible client ports. The scan is as follows: the attacker sends a SYN to the server spoofed as if from the client; if there is already a connection through the client port specified in the SYN packet, then the server ignores the spoofed SYN. Otherwise the server sends a SYN/ACK packet to the client who will respond in RST. The attacker uses the global IP-ID to test whether the client sent a packet in response.

Implementing this method for probing the client port has a few challenges: (a) this technique is filtered by typical client firewalls (e.g., Windows Firewall) that will discard the SYN/ACK server response in case that the client did not first send a SYN. (b) attacker must run a synchronized attack, querying for the client IP-ID, then assume that the server probe had arrived and query for the IP-ID again; if during this time \wini\ sends a packet or server SYN/ACK does not yet arrive then the test is invalid.

In contrast, we create the connection using the puppet and identify the client port by using an insight on Windows port allocation paradigm. This allows us to form a connection with an `interesting' server and efficiently detect its parameters (see Section \ref{exploiting:S:conn}).

Second, the attacker in \cite{lkm:phrack:07} must cope ongoing traffic over the victim connection itself. Such traffic fails the binary search for the client sequence number (see Section \ref{Injection:AckEsposure}) since this phase requires specifying a valid sequence number (which keeps changing due to traffic on the connection). Moreover, \cite{lkm:phrack:07} does not describe how to implement the queries to (1) avoid firewall filtering and (2) detect network losses. In the approach presented in this paper, the attacker controls the connection since her puppet makes the requests for the server. Hence she is able to avoid traffic on the connection while exposing the sequence numbers. The legitimate TCP connection with the client is used to implement the queries (see details in Section \ref{Injection:Experiments}).

\begin{figure}[htb]
	\centering
		\includegraphics[width=0.42\textwidth]{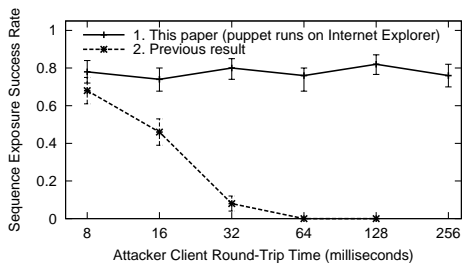}
	\caption{Comparison of sequence exposure techniques. Each measurement is the average of 50 runs, error bars mark standard deviations.}
	\label{fig:phrackcomparison}
\end{figure}

In Figure \ref{fig:phrackcomparison} we compare the success rates of our attack to that described in \cite{lkm:phrack:07} where the victim connection in while running the attack in \cite{lkm:phrack:07} has only a modest 10 kbps traffic rate (since attacker does not control the traffic rate in this case). The comparison is for different network delays between the client and attacker; the longer the delay, the more time until the attacker receives feedback and the more traffic that passes on the connection. Since \cite{lkm:phrack:07} does not specify how to implement the queries, we used our method, i.e., on a TCP connection between the client and the attacker. We assume that the attacker in \cite{lkm:phrack:07} successfully detects the client port (despite the challenges above). We also assume that the client sends an average of 32 packets per second to other peers. 

The third difference between our approach to \cite{lkm:phrack:07} regards to the practical challenge of performing a `meaningful' injection. That is, after a successful exposure of sequence numbers, the attacker should identify the right time to inject his data; For example, to perform the XSS attack, the spoofed response must arrive after the client had sent a request; it is hard for an off-path attacker to detect that time. In contrast, the attacks in this initiate the request using the puppet and inject the response (see Section \ref{exploiting:S}).

\end{document}